\documentclass[twocolumn]{aastex62}
\usepackage{bm}
\usepackage{graphicx}
\usepackage{wasysym}
\begin{document}

\title{\textbf{\Large{Dynamics of multiple bodies in a corotation resonance: \\Conserved quantities and relevance to ring arcs}}}

\author[0000-0001-7522-7806]{Joseph A. A'Hearn}
\affiliation{Department of Physics, University of Idaho, Moscow, Idaho, USA}

\author[0000-0002-8592-0812]{Matthew M. Hedman}
\affiliation{Department of Physics, University of Idaho, Moscow, Idaho, USA}

\author[0000-0002-4416-8011]{Maryame El Moutamid}
\affiliation{Cornell Center for Astrophysics and Planetary Science, Cornell University, Ithaca, New York, USA}
\affiliation{Carl Sagan Institute, Cornell University, Ithaca, New York, USA}

\correspondingauthor{Joseph A'Hearn}
\email{jahearn@uidaho.edu}

\begin{abstract}
The interactions among objects in a mean motion resonance are important for the orbital evolution of satellites and rings, 
especially Saturn's ring arcs and associated moons. 
In this work, we examine interactions among massive bodies 
in the same corotation eccentricity resonance site that affect the orbital evolution of those bodies using numerical simulations.
During these simulations, the bodies exchange angular momentum and energy during close encounters, altering their orbits. 
This energy exchange, however, does not mean that one body necessarily moves closer to exact corotation when the other moves away from it. 
Indeed, if one object moves towards one of these sites, the other object is equally likely to move towards or away from it.
This happens because the timescale of these close encounters is short compared to the synodic period between these particles and the secondary mass 
(i.e., the timescale where corotation sites can be treated as potential maxima).
Because the timescale of a gravitational encounter is comparable to the timescale of a collision, 
we could expect energy to be exchanged in a similar way for collisional interactions. 
In that case, these findings could be relevant for denser systems like the arcs in Neptune's Adams ring 
and how they can be maintained in the face of frequent inelastic collisions.
\end{abstract}


\section{Introduction \label{intro}}
\begin{figure}
\includegraphics[width=\linewidth]{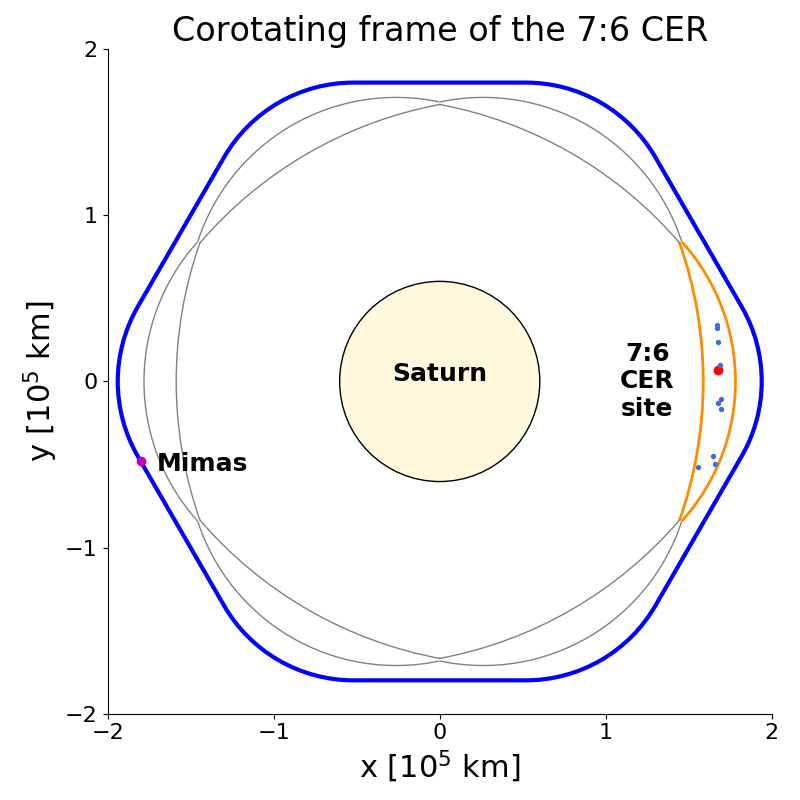}
\caption{
The geometry of a corotation eccentricity resonance. 
In the corotating frame of each of six fixed points of the 7:6 corotation eccentricity resonance with Mimas, 
Mimas traces out a rounded hexagonal shape (blue). 
This creates six corotation sites where material can become trapped (within the gray boundaries).
Aegaeon's ring arc consists of trapped material in just one of the six corotation sites (within the orange boundary). 
For illustration purposes, we have exaggerated the eccentricity of Mimas by about a factor of 2
and we have stretched the radial boundaries of the corotation sites 
as well as the distances the ring arc particles appear from the semi-major axis of the fixed point (using a nonlinear function). 
The real width and length of the corotation sites are defined in Equations \ref{cer_width} and \ref{cer_length}.
The positions of Mimas (magenta) and the ring arc bodies (royal blue for the equal-mass bodies and red for the larger-mass body) 
depicted here are from their initial positions in one of our simulations (see Section \ref{methods}). 
The phase space plots in other figures are parametric projections of the single corotation site with the ring arc bodies. 
\label{corotating_frame}}
\end{figure}

In our solar system, both Saturn and Neptune have ring arcs.
Saturn's ring arcs are confined longitudinally due to corotation eccentricity resonances with Saturn's moon Mimas:
Aegaeon and its ring arc are in a 7:6 corotation resonance with Mimas \citep{2007Sci...317..653H, 2010Icar..207..433H} (see Figure \ref{corotating_frame}),
Anthe and its ring arc are in a 10:11 corotation resonance with Mimas \citep{2008Icar..195..765C, 2009Icar..199..378H}, while
Methone and its ring arc are in a 14:15 corotation resonance with Mimas \citep{2006AJ....132..692S, 2009Icar..199..378H}. 
Neptune's ring arcs are also confined longitudinally \citep{1989Sci...246.1422S} and move at rates close to a 42:43 corotation resonance with Galatea, 
which may be the explanation for their confinement \citep{1986AJ.....92..490G, 1991Sci...253..995P, 2002Natur.417...45N}.
Deviations from this exact rate, however, could support other explanations,
like the presence of undetected co-orbital satellites \citep{1998Sci...282.1102S, 2014A&amp;A...563A.133R}
or a three-body resonance with Galatea and Larissa \citep{2017DPS....4910401S}.

The interactions among objects in a mean motion resonance are important for the orbital evolution of satellites and rings. 
For example, multiple authors have looked at the importance of the nearby Lindblad resonance in maintaining energy among ring arc particles, 
since we would otherwise expect energy to dissipate due to collisions \citep{1986AJ.....92..490G, 1991Sci...253..995P, 2002Natur.417...45N}. 
Thus far, however, there have not been detailed investigations of interactions of multiple bodies within a corotation resonance. 
Recent work has focused instead on the motions of individual objects in these resonances. For example, 
\citet{2014CeMDA.118..235E} developed the CoraLin model providing a description of a time-averaged Hamiltonian of the three-body system
and showed that intermediate distances between the corotation and Lindblad resonances yield a region of chaotic motion. 
\citet{2017MNRAS.469.2380E} then studied the capture of massless particles into corotation eccentricity resonances.
\citet{2017MNRAS.470.3750M} performed a study on the long-term evolution ($10^5$ years) of Saturn's moons Aegaeon, Methone, Anthe, and Pallene.
\citet{2017Icar..284..206S} looked at the dynamics of small particles in corotation resonances with Anthe and Methone. 
\citet{2018MNRAS.475.5474M} examined the influence of Aegaeon on $\mu$m-sized dust particles, acknowledging that larger particles (cm- to m-sized) could also be present.
Although these last studies have discussed satellite perturbations on dust, 
no study has explored mutual interactions between ring arc bodies. 
In this work, we consider how interactions among massive bodies 
in the same corotation eccentricity resonance site affect their orbital evolution on short timescales.
We find that the time-averaged Hamiltonian that is so useful for describing three-body motion is no longer appropriate for mutual encounters, 
which has implications for the stability of arcs where such interactions are common.  

In Section \ref{bg}, we cover the background of the three-body problem where the third body is in a corotation eccentricity resonance with the secondary body. 
In Section \ref{methods}, we describe how we use numerical simulations to investigate the interactions between two or more massive bodies trapped in the same corotation resonance. 
In Section \ref{results}, we consider what happens when two or more massive bodies share the same corotation resonance site. 
In Section \ref{disc}, we describe the results of these simulations, which demonstrate that the time-averaged energy defining the corotation resonance is not conserved. 

\section{Background \label{bg}}
In this section, we review mean motion resonances around an oblate primary, body 1,
and then discuss the dynamics of the three-body problem 
where the secondary, body 2, holds a third body, body 3, in a corotation eccentricity resonance.
We assume a hierarchical system in which $M_1 \gg M_2 \gg M_3$.

Mean motion resonances occur when the orbital motions of two objects in orbit around a primary body are commensurate with each other. 
For objects in orbit around giant planets, however, the planet's oblateness splits each resonance into multiple sub-resonances of different types. 
For a test particle orbiting around an oblate central mass, the gravitational potential is 
\begin{equation}
V = -\frac{GM_1}{r}\left[ 1-\sum_{\tiny{i=1}}^{\infty} J_{2i} \left( \frac{R}{r} \right)^{2i} P_{2i} \left(\textrm{sin} \alpha \right) \right]
\label{V_Sat}
\end{equation} 
where $G$ is the gravitational constant, $M_1$ is the mass of the primary body, 
$r$ is the distance between the test particle and the center of $M_1$, 
the $J_{2i}$ terms are zonal gravity harmonic coefficients,
and the $P_{2i}$ terms are Legendre polynomials in sin$\alpha$, where the angle $\alpha$ is measured from the equatorial plane of the primary body.
These terms in the potential alter the expressions for the particle's mean motion $n$ and radial epicyclic frequency $\kappa$ \citep{1999ssd..book.....M, 2006CeMDA..94..237R}:
\begin{equation}
\footnotesize{n^2 \simeq \frac{GM}{a^3}\left[ 1+\frac{3}{2}J_2\left( \frac{R}{a} \right)^2-\frac{15}{8}J_4\left( \frac{R}{a} \right)^4+\frac{35}{16}J_6\left( \frac{R}{a} \right)^6 \right]}
\end{equation} 
\begin{equation}
\footnotesize{\kappa^2 \simeq \frac{GM}{a^3}\left[ 1-\frac{3}{2}J_2\left( \frac{R}{a} \right)^2+\frac{45}{8}J_4\left( \frac{R}{a} \right)^4-\frac{175}{16}J_6\left( \frac{R}{a} \right)^6 \right]}
\end{equation} 
where $a$ is the semi-major axis of the test particle. 
These extra terms in $n$ and $\kappa$ cause the locations of resonances with a secondary mass to split. 
Specifically, for any integer $j$, a Lindblad eccentricity resonance occurs where 
\begin{equation}
j n_{\textrm{\tiny{LER}}} = \left(j+1\right)n_2 - \dot{\varpi}_{\textrm{\tiny{LER}}} 
\end{equation} 
while the corresponding corotation eccentricity resonance occurs where 
\begin{equation}
j n_{\textrm{\tiny{CER}}} = \left(j+1\right)n_2 - \dot{\varpi}_2 
\label{cer_eq}
\end{equation} 
since the pericenter precession rate is given by
\begin{equation}
\dot{\varpi} = n - \kappa
\end{equation} 



Corotation eccentricity resonances exist when the perturbing body has non-zero eccentricity. 
The eccentricity of Mimas, $e = 0.0196$ \citep{2010DDA....41.0805J}, for example, 
is enough to provide large corotation sites for Aegaeon, Anthe, and Methone. 
The main effect of the corotation resonance is to drive oscillations in the perturbed body's semi-major axis and mean longitude  
around a series of points where Equation \ref{cer_eq} is exactly satisfied. 
These points correspond to $j$ equally-spaced corotating longitudes $\lambda_{\textrm{\tiny{CER}}}$ at the same semi-major axis $a_{\textrm{\tiny{CER}}}$.
 
\begin{figure}
\includegraphics[width=\linewidth]{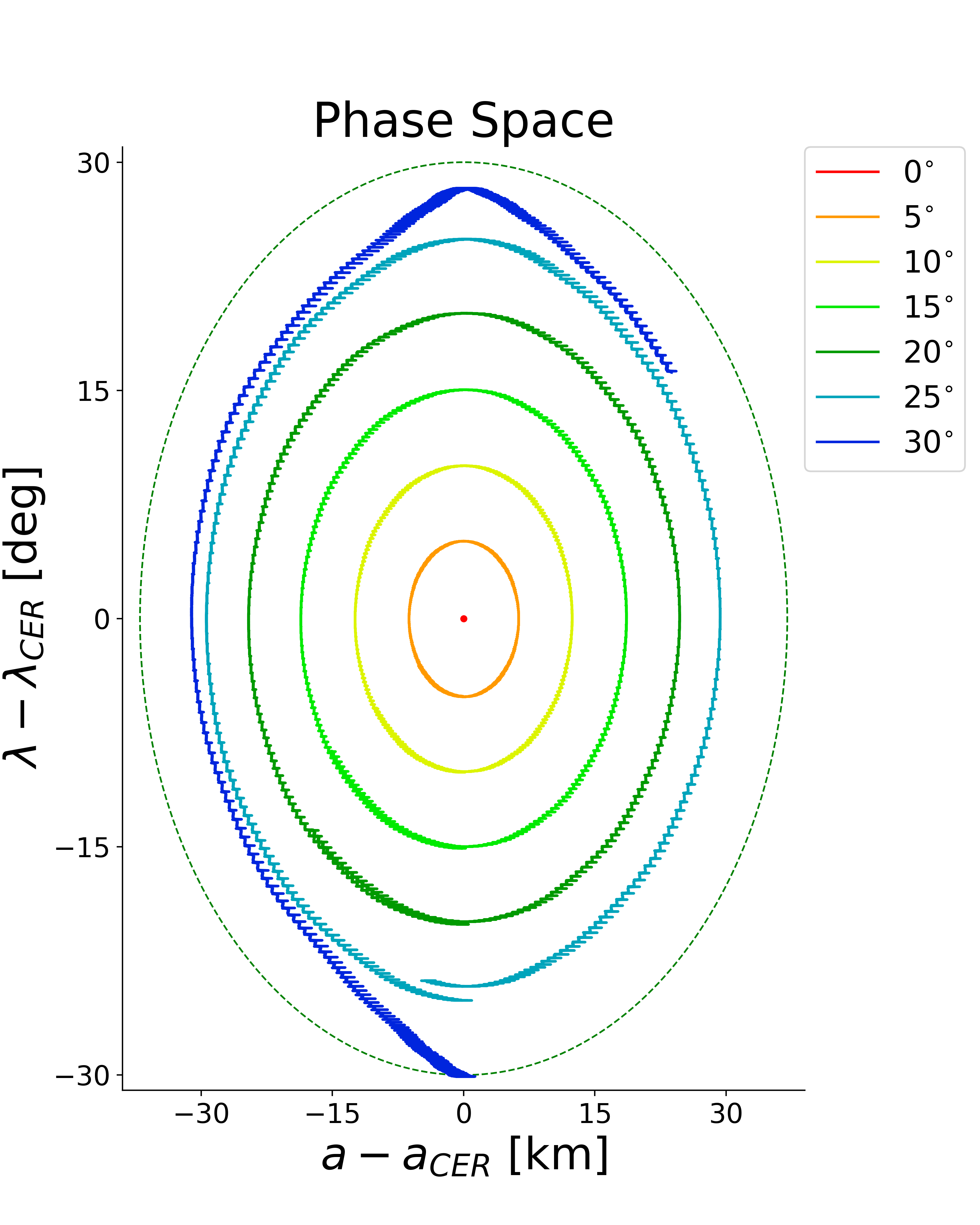}
\caption{
A body in a corotation resonance traces out a quasi-elliptical path in phase space, moving in the clockwise direction.
The paths traced out in phase space seen here are from separate 4-year simulations in which the third body
is initially placed 0, 5, 10, 15, 20, 25, and then 30 degrees of mean longitude behind the exact corotating longitude, 
and at $a_{\textrm{\tiny{CER}}} = 167506.5$ km. 
The dashed ellipse marks an approximate boundary for the corotation resonance,
though it can be seen that near the fringe of the resonance site bodies trace out paths that resemble an American football rather than an ellipse. 
The curves do not quite close on themselves at large distances from the resonance center due to additional perturbations from the Lindblad resonance, 
which is located at about +19 km.
\label{contours}}
\end{figure}

It is useful to depict this motion as well as a body's location with a phase space of corotating longitude vs. semi-major axis, as shown in Figure \ref{contours}.
In this phase space, bodies in the corotation resonance follow quasi-elliptical trajectories. 
The center of the ``ellipse" is the phase space location of exact corotation resonance.
This is a local potential maximum in a field that is time-averaged over the synodic period between the secondary and tertiary bodies
\citep{1986AJ.....92..490G, 1991Icar...89..197S, 1995netr.conf..703P, 2002Natur.417...45N}.

The width of a corotation eccentricity resonance (horizontal distance in Figure \ref{contours}), for sufficiently large $j$, can be approximated as
\citep{2014CeMDA.118..235E}
\begin{equation}
W_{\textrm{\tiny{CER}}} \simeq 4.136 \sqrt{|j| e_2 \frac{M_2}{M_1} \frac{a_{\textrm{\tiny{CER}}}^3}{a_2}}
\label{cer_width}
\end{equation} 
where $a_{\textrm{\tiny{CER}}}$ is the semi-major axis of the corotation eccentricity resonance, 
$a_2$ is the semi-major axis of the secondary body, $e_2$ is the eccentricity of the secondary,
$M_2$ is the mass of the secondary, $M_1$ is the mass of the primary (central) body, 
and the coefficient $4.136$ has absorbed constants as well as combinations of Laplace coefficients \citep{1961mcm..book.....B}.

The length of the corotation eccentricity resonance (vertical distance in Figure \ref{contours}), measured in degrees, is simply
\begin{equation}
L_{\textrm{\tiny{CER}}} = \frac{360^{\circ}}{|j|} 
\label{cer_length}
\end{equation} 
for an inner resonance (replace $|j|$ with $|j+1|$ for an outer resonance).
From the width and length, the resulting boundary of the corotation resonance can then be approximated as an ellipse.
We can calculate a normalized distance $s$ from exact resonance in phase space to help our analysis. 
\begin{equation}
\label{eq-s}
s^2 = \left( \frac{a - a_{\textrm{\tiny{CER}}}}{0.62 W_{\textrm{\tiny{CER}}}} \right)^2 + \left( \frac{\lambda - \lambda_{\textrm{\tiny{CER}}}}{0.5 L_{\textrm{\tiny{CER}}}} \right)^2
\end{equation} 
where $s < 1$ for bodies in the corotation resonance and $s > 1$ for bodies outside the corotation resonance. 
The coefficients in the denominators apply to the 7:6 corotation resonance and are empirically determined such that 
the phase space distance $s$ remains relatively constant over the course of librations in the corotation resonance. 


\clearpage
\section{Methods \label{methods}}

To examine the dynamics of multiple bodies in a corotation eccentricity resonance, 
we numerically simulated the motion of objects with orbits similar to Aegaeon, 
which orbits Saturn within the 7:6 CER created by Mimas.
For orbital simulations, we used Mercury6 code \citep{1999MNRAS.304..793C}. 
Our orbital simulations considered Saturn as the central mass and included terms up to $J_6$ in its gravitational field. 
The constants used for these simulations were taken from \citet{2006AJ....132.2520J} and are found in Table \ref{tbl-1}. 
\begin{table}
\begin{center}
\caption{Parameters of Saturn used for numerical simulations, from \citet{2006AJ....132.2520J} \label{tbl-1}}
\begin{tabular}{crr}
\tableline
\tableline
Parameter & Value \\
\tableline
R$_{\saturn}$ &60268 km \\
GM$_{\saturn}$ &37931207.7 km$^3$s$^{-2}$\\
$J_2$ &1.629071$\times 10^{-2}$\\
$J_4$ &-9.3583$\times 10^{-4}$ \\
$J_6$ &8.614$\times 10^{-5}$ \\
\tableline
\end{tabular}
\end{center}
\end{table}
Mimas was included in all simulations, with its initial state vectors from an arbitrary date (UTC 2010-100T00:00:00) 
according to the SPICE kernel sat393.bsp (\citealt{1996P&amp;SS...44...65A}, Table \ref{tbl-2}).

\begin{table}
\begin{center}
\caption{Parameters of Mimas used for numerical simulations, 
corresponding to its position at UTC 2010-100T00:00:00 \label{tbl-2}}
\begin{tabular}{cr}
\tableline
\tableline
Parameter & Value \\
\tableline
M &6.597$\times 10^{-8}$M$_{\saturn}$\\
$x$ &8.81807403961$\times 10^{-4}$ AU\\
$y$ &8.80075627975$\times 10^{-4}$ AU\\
$z$ &1.2509303717$\times 10^{-5}$ AU\\
$\dot{x}$ &-5.70335493828$\times 10^{-3}$ AU/day\\
$\dot{y}$ &5.93108634437$\times 10^{-3}$ AU/day\\
$\dot{z}$ &2.10785127451$\times 10^{-4}$ AU/day\\
\tableline
\end{tabular}
\end{center}
\end{table}

For each simulation, we modified the initial state vectors of masses we placed in the same 7:6 corotation resonance site with Mimas
to set up a system similar to Aegaeon's ring arc.  
We used time-steps of 0.01 days in order to observe carefully what happens during a close encounter. 
After an initial test with the Bulirsch-Stoer integrator, we chose the hybrid symplectic/Bulirsch-Stoer integrator with the changeover at 3 Hill radii.
Since a libration in the phase space of the 7:6 corotation resonance with Mimas takes about four years \citep{2010Icar..207..433H, 2017MNRAS.470.3750M}, 
and a close encounter between any two bodies generally occurs twice each libration,
ten-year simulations were sufficient to observe an average of 15 close encounters per body per simulation. 

Aegaeon's ellipsoidal axes and inferred mean density \citep{2013Icar..226..999T} give a mass estimate of $1.0 \times 10^{11}$ kg, 
but encounters with such small masses produce very small changes in the phase space distance $s$. 
Hence, in order to better document the changes in the particles' orbits during a close encounter, 
we consider much larger mass objects: 10 objects with masses of $2 \times 10^{13}$ kg and (optionally) one object with a mass of $6 \times 10^{14}$ kg.
This preserves the estimated mass ratio of $0.3$ of the total mass of all the other bodies in Aegaeon's ring arc to the mass of Aegaeon \citep{2010Icar..207..433H}.
We verified that increasing the masses changes the results only quantitatively, not qualitatively, 
by experimenting with smaller masses over larger timescales and 
by observing that changes in semi-major axis in asymmetric mass interactions scale linearly with mass.

The initial positions of all bodies were distributed randomly in $\left(e \textrm{cos} \varpi, e \textrm{sin} \varpi \right)$ space with $0.00001 < e < 0.0008$
and $\left(I \textrm{cos} \Omega, I \textrm{sin} \Omega \right)$ space with $0.00001^{\circ} < I < 0.0917^{\circ}$.
For every simulation with the more massive body, an additional simulation without the more massive body was performed, 
but with the same initial positions and velocities given to the other bodies. 

After running each orbital simulation, we converted the state vectors at each time-step 
to geometric orbital elements using the equations and iterative method found in \cite{2006CeMDA..94..237R}. 
Using the geometric elements is important when considering orbits around an oblate planet like Saturn 
because they do not exhibit the artificial orbit-period variations seen in osculating elements.

\section{Results \label{results}}
\begin{figure*}[hbtp]
\includegraphics[width=\textwidth]{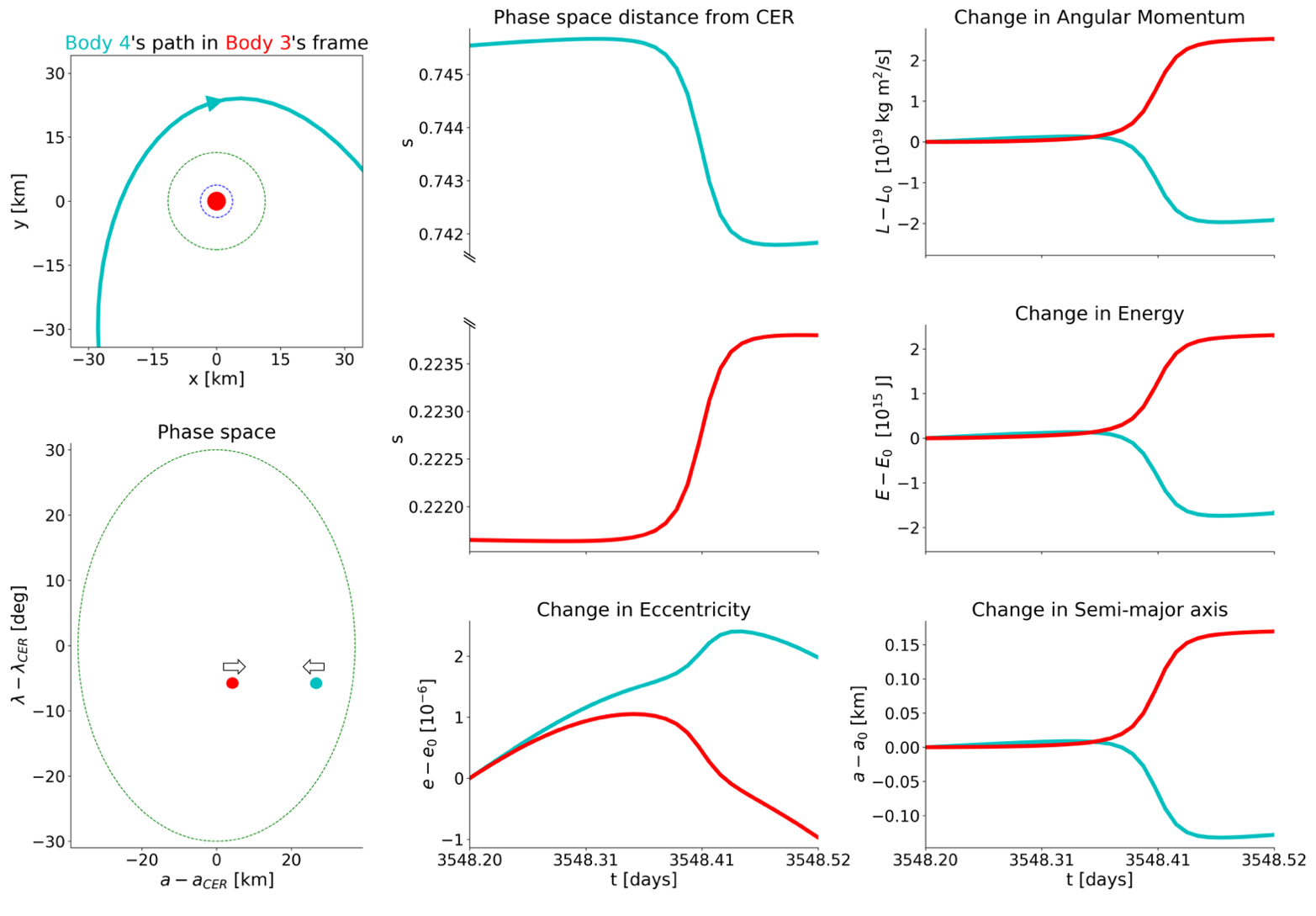}
\caption{
A sample close encounter of equal-mass bodies. 
The result of this close encounter is that one body (cyan) moves closer to exact corotation while the other (red) moves away from exact corotation. 
The mass of each body in this close encounter is $2 \times 10^{13}$ kg. 
The eccentricity at the start of the close encounter, $e_0$, was $8.078 \times 10^{-4}$ for body 3 and $6.779 \times 10^{-4}$ for body 4.  
The angular momentum at the start of the close encounter, $L_0$, was $5.0217 \times 10^{25}$ kg m$^2$/s for body 3 and $5.0221 \times 10^{25}$ kg m$^2$/s for body 4. 
The total energy at the start of the close encounter, $E_0$, was $-2.2495 \times 10^{21}$ J for body 3 and $-2.2492 \times 10^{21}$ J for body 4.
The semi-major axis at the start of the close encounter, $a_0$, was $167511$ km for body 3 and $167534$ km for body 4.
}
\label{fig-one_in_one_out}
\end{figure*}

If we add another body of comparable mass (i.e., $M_3 \simeq M_4$) to the same corotation site, the bodies in the corotation site,
which we call body 3 and body 4, can have close encounters with each other. 
Although they will have slightly different semi-major axes, the forcing due to the nearby Lindblad resonance 
provides enough eccentricity for the bodies to have a close encounter when they share the same mean longitude,
which generally happens twice per libration period. 
We describe how we quantify these interactions in Section \ref{qi}, and summarize the distribution of outcomes in Section \ref{probability}.

\subsection{Quantifying close encounters \label{qi}}

We define a close encounter as any time one body is less than 6 Hill radii from another. 
For each close encounter, we generate a series of plots documenting change in the orbital properties of the two bodies 
(see Figure \ref{fig-one_in_one_out}). These plots show the path of one body in the fixed frame of the other, 
the bodies' trajectories in phase space during the interaction,
their phase space distance from exact corotation defined in Equation \ref{eq-s}, 
and their changes in semi-major axis, eccentricity, angular momentum, and energy over the course of the interaction. 

The top left panel of these plots shows the encounter in the frame where body 3 is fixed, with up being the direction of orbital motion. 
The dotted blue circle around the red body marks its Hill radius. 
The dotted green circle around the red body marks 3 Hill radii, 
which is the boundary inside of which the integrator uses the Bulirsch-Stoer algorithm. 
The cyan path shows the motion traced out by body 4, with an arrow indicating the direction of motion. 

The next plot down shows the paths in phase space of body 3 (red) and body 4 (cyan) during the close encounter. 
Again, the bodies can interact even though they are at different semi-major axes because of their orbital eccentricities. 
In most cases, the change in semi-major axis during a close encounter is small compared to the oscillations of semi-major axis over the course of a libration period. 

\begin{figure*}[hbtp]
\includegraphics[width=\textwidth]{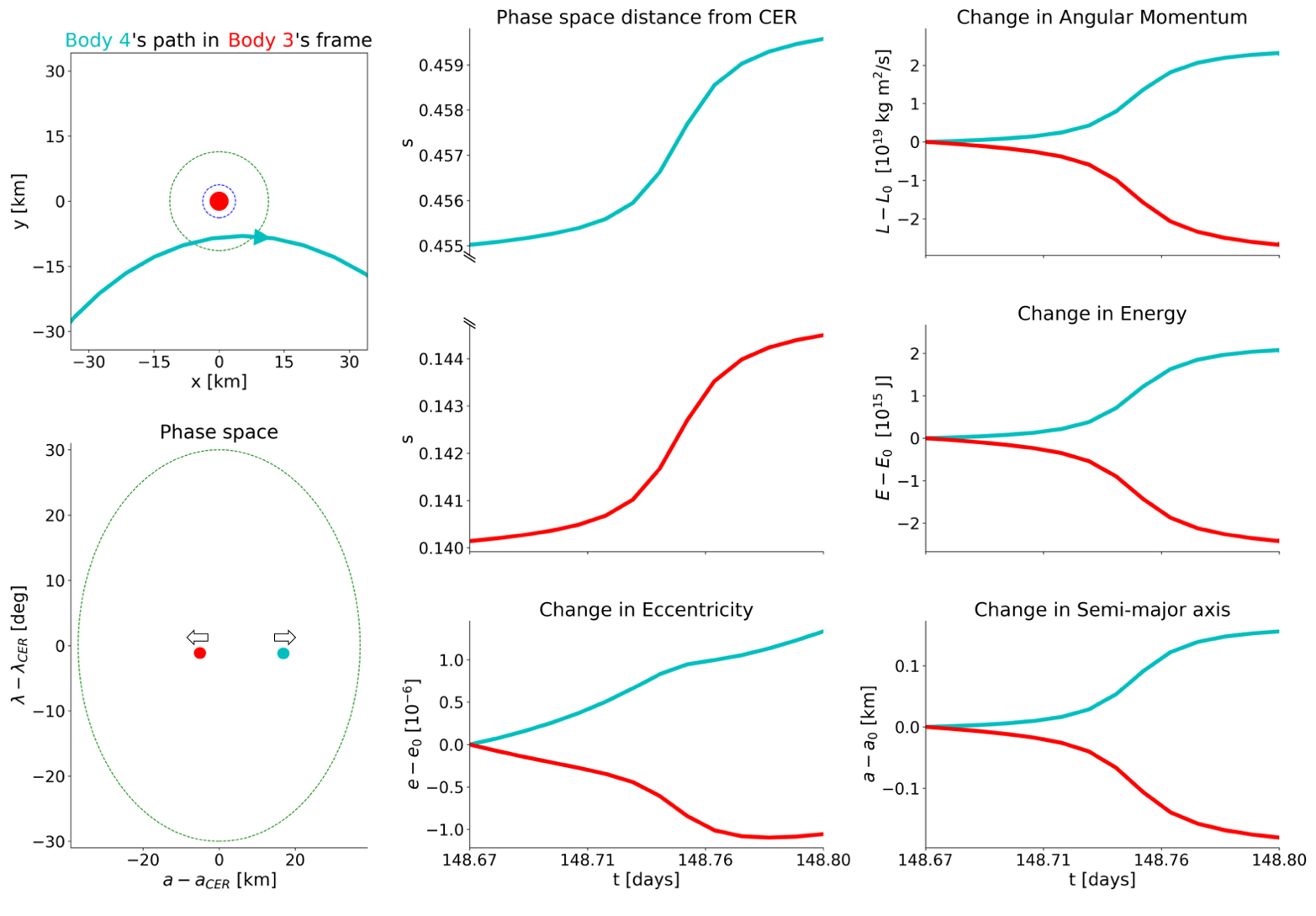}
\caption{
Close encounter of symmetric masses in which both bodies move away from exact corotation. 
The mass of each body in this close encounter is $2 \times 10^{13}$ kg. 
The eccentricity at the start of the close encounter, $e_0$, was $2.418 \times 10^{-4}$ for body 3 and $6.32 \times 10^{-4}$ for body 4.  
The angular momentum at the start of the close encounter, $L_0$, was $5.0216 \times 10^{25}$ kg m$^2$/s for body 3 and $5.0219 \times 10^{25}$ kg m$^2$/s for body 4. 
The total energy at the start of the close encounter, $E_0$, was $-2.2496 \times 10^{21}$ J for body 3 and $-2.2493 \times 10^{21}$ J for body 4.
The semi-major axis at the start of the close encounter, $a_0$, was $167502$ km for body 3 and $167523$ km for body 4.
}
\label{fig-both_out}
\end{figure*}

The first two plots of the second column show the evolution over time of the phase space distance from exact corotation.
The top one (cyan) is for body 4; the bottom one (red) is for body 3.
For the interaction shown in Figure \ref{fig-one_in_one_out}, body 3 (red) moves away from exact resonance while body 4 (cyan) moves towards exact resonance.

The bottom plot in the second column shows the change over time in eccentricity.
The plots in the third column show the change over time in angular momentum, energy, and semi-major axis.
An equation for angular momentum in the context of the two-body problem (cf. Equation 2.26 in \citealt{1999ssd..book.....M})
\begin{equation}
L = m \sqrt{GMa(1-e^2)}
\end{equation} 
shows that angular momentum is proportional to the square root of the semi-major axis. 
This relation is reflected in the plots shown. 
When we compute angular momentum, however, we use a more fundamental equation (cf. Equations 2.128-129 in \citealt{1999ssd..book.....M}):
\begin{equation}
L = m |\vec{r} \times \vec{v}|
\end{equation} 
Total energy is conserved during these close encounters as long as it is computed in the fundamental way:
\begin{equation}
E = \sum_{\tiny{i}}^{N} \left(V_{1,i} + \frac{1}{2} m_i v_i^2 \right) - \sum_{\tiny{i \neq j}}^{N} \frac{G m_i m_j}{r_{ij}} 
\end{equation} 
where $V_{1,i}$ is the potential energy due to the oblate central body as computed in Equation \ref{V_Sat}. 
Note in Figure \ref{fig-one_in_one_out} the changes in $L$ and $E$ have opposite signs for bodies 3 and 4, 
consistent with the two objects exchanging energy and angular momentum.

\subsection{Probability of different outcomes \label{probability}}
\begin{figure*}[hbtp]
\includegraphics[width=\textwidth]{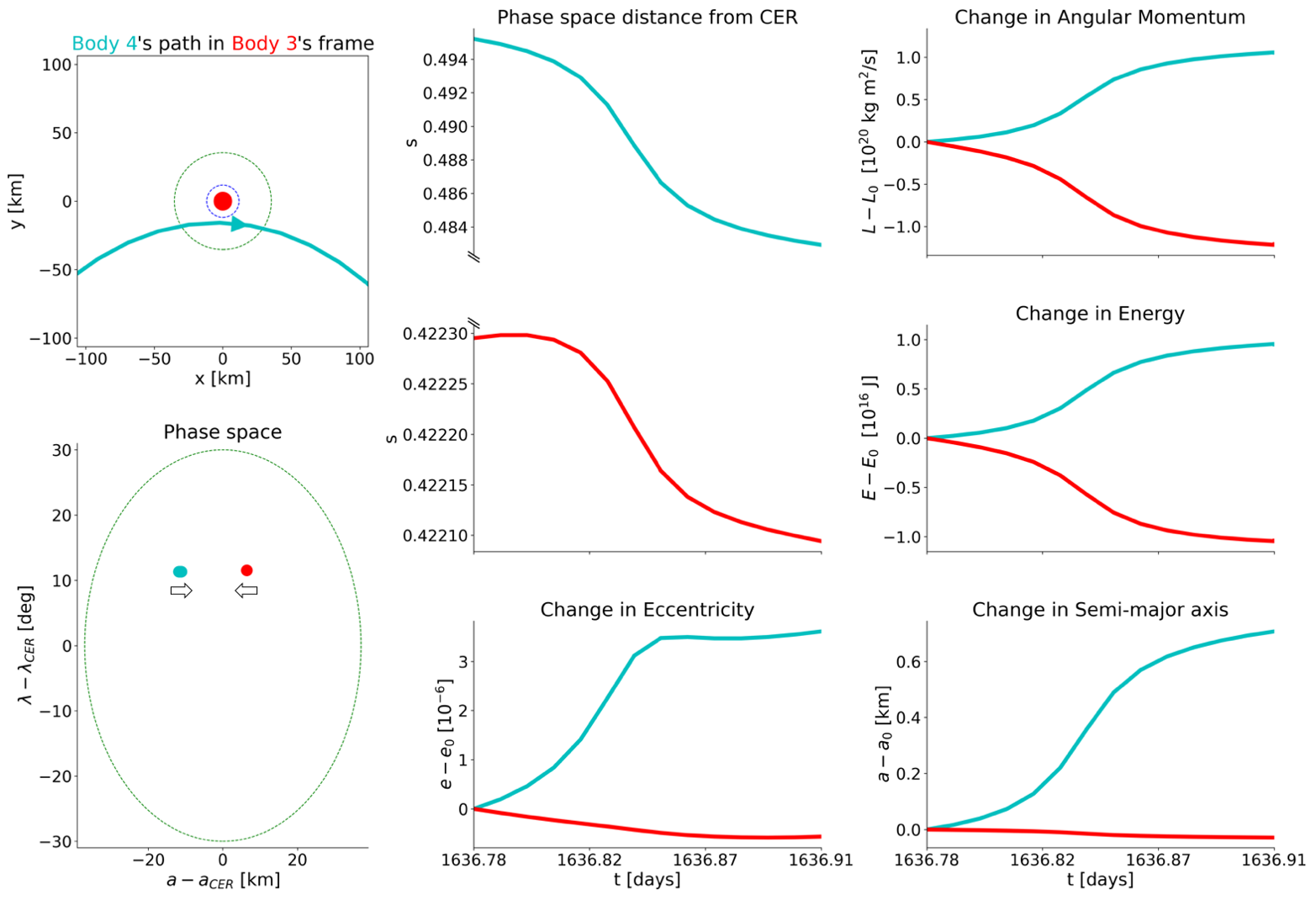}
\caption{
Close encounter of asymmetric masses in which both bodies move towards exact corotation.
The mass of body 3 (red) is $6 \times 10^{14}$ kg and the mass of body 4 (cyan) is $2 \times 10^{13}$ kg. 
The eccentricity at the start of the close encounter, $e_0$, was $1.1134 \times 10^{-3}$ for body 3 and $1.6571 \times 10^{-3}$ for body 4.  
The angular momentum at the start of the close encounter, $L_0$, was $1.5065 \times 10^{27}$ kg m$^2$/s for body 3 and $5.0215 \times 10^{25}$ kg m$^2$/s for body 4. 
The total energy at the start of the close encounter, $E_0$, was $-6.7484 \times 10^{22}$ J for body 3 and $-2.2497 \times 10^{21}$ J for body 4.
The semi-major axis at the start of the close encounter, $a_0$, was $167513$ km for body 3 and $167495$ km for body 4.
}
\label{fig-both_in}
\end{figure*}
Given the standard three-body picture of the co-rotation resonance, 
where the sites of exact corotation are treated as (time-averaged) potential maxima, 
one might expect that conservation of energy would require that if the phase space distance $s$ of one body decreases, that of the other body must increase.
In fact, however, we find encounters are equally likely to cause objects to both move in the same direction relative to the resonance center.
From our simulations, about 5000 close encounters have been analyzed. 
In $49\%$ of cases, one body moves closer to exact resonance while the other moves away, like the interaction depicted in Figure \ref{fig-one_in_one_out}. 
In $26\%$ of cases, both bodies move away from exact resonance. 
In $25\%$ of cases, both bodies move towards exact resonance. 
(With 5000 events the statistical uncertainties in these fractions are all 2\%.)
Examples of these encounters are shown in Figures \ref{fig-both_out} and \ref{fig-both_in}.
These encounters explicitly conserve energy and angular momentum, 
so these unexpected results are not due to an error in the code. 
The conservation of energy and angular momentum holds even in cases where the bodies have unequal masses, such as the encounter shown in Figure \ref{fig-both_in}.

\begin{figure*}[hbtp]
\includegraphics[width=\textwidth]{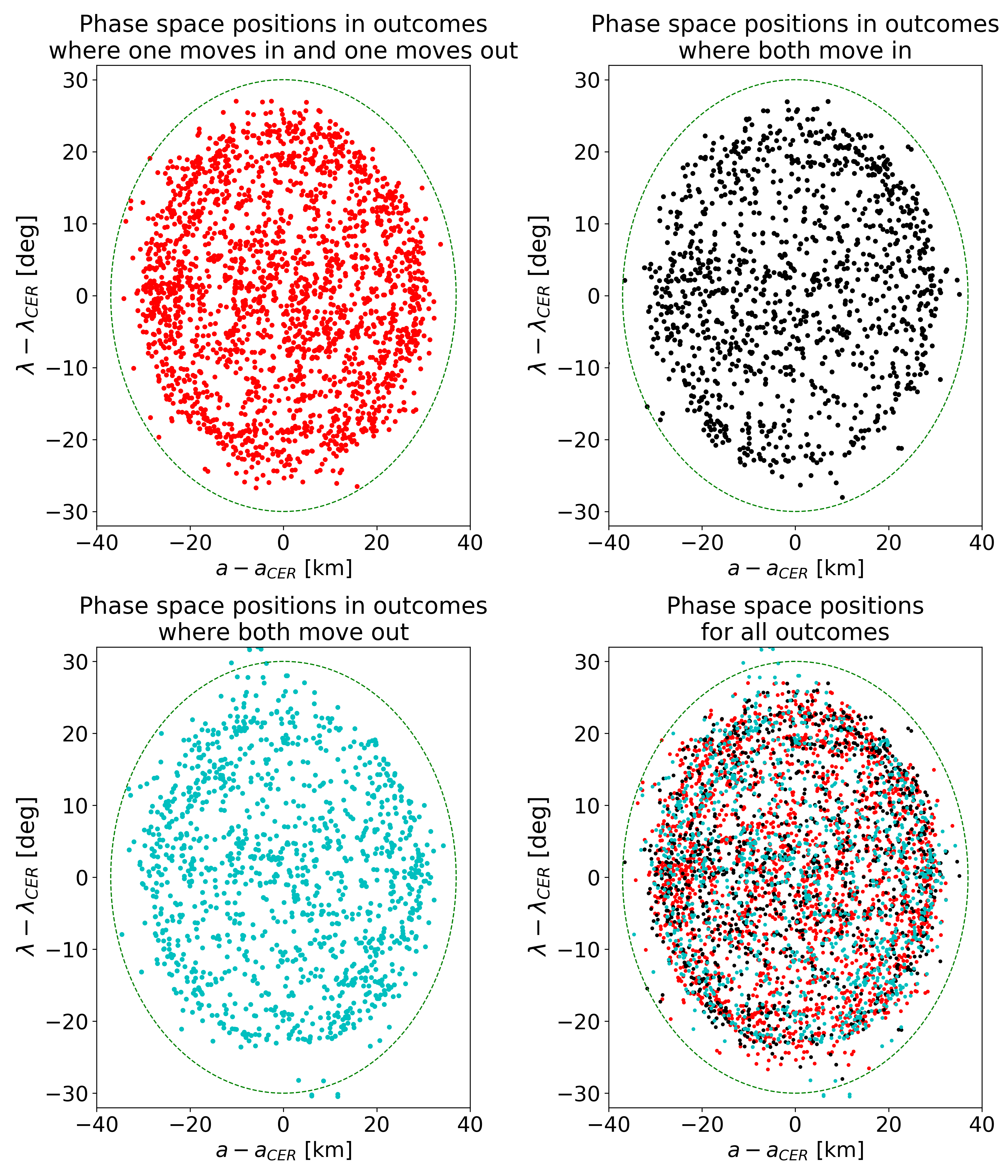}
\caption{
Positions in phase space of each body at the beginning of an encounter, 
organized by the outcome of the encounter. 
The uniformity in these distributions is evidence that 
there is no strong trend for a certain outcome based on where the body is in phase space. 
The randomness in outcomes throughout phase space suggests that 
the long-term evolution of ring arc particles does not differ qualitatively from the evolution 
observed in these ten-year simulations.
}
\label{all_encounters}
\end{figure*}

We investigated whether any regions of phase space favored certain outcomes. 
No obvious dependence on any specific part of phase space for a certain type of encounter outcome to occur can be seen in Figure \ref{all_encounters}. 
This randomness in outcomes suggests that the long-term evolution of bodies in a corotation eccentricity resonance 
does not differ qualitatively from the evolution observed in these ten-year simulations.

\section{Discussion \label{disc}}

The above simulations clearly show that close encounters between bodies within a corotation resonance do not conserve the phase space distance $s$. 
To understand why this is the case, we first examine the individual encounters shown in Figures \ref{fig-one_in_one_out}-\ref{fig-both_in}, 
and show that the changes in semi-major axis are consistent with the encounter geometries. 
We then argue that the classical understanding of energy surfaces in corotation resonances is not applicable here because the encounters occur on very short timescales. 
Finally, we highlight some potential implications of these findings for the stability of ring arcs.

In the close encounter shown in Figure \ref{fig-one_in_one_out}, body 4 (cyan) passes by ahead of body 3 (red) in their direction of orbital motion (up). 
Because of this, angular momentum and energy are transferred from body 4 to body 3. 
This determines which direction the bodies move in phase space. 
Since at the beginning of the interaction body 3 has a semi-major axis greater than the semi-major axis of exact corotation, 
and then gains angular momentum and energy, its semi-major axis increases, and it thus moves away from exact corotation. 
Body 4 also begins the interaction with a semi-major axis greater than the semi-major axis of exact corotation, 
but because it loses angular momentum and energy, its semi-major axis decreases, so it moves towards exact corotation.  

In the close encounters shown in Figure \ref{fig-both_out} and in Figure \ref{fig-both_in}, body 4 (cyan) passes by behind body 3 (red), 
so angular momentum and energy are transferred from body 3 to body 4.  
In Figure \ref{fig-both_out}, body 3 begins the interaction with a semi-major axis less than the semi-major axis of exact corotation, 
whereas body 4 begins with a semi-major axis greater than that of exact corotation. 
Since body 3 is losing angular momentum and energy, its semi-major axis decreases, and it moves away from exact corotation. 
Since body 4 is gaining angular momentum and energy, its semi-major axis increases, so it also moves away from exact corotation. 

In Figure \ref{fig-both_in}, body 3 begins the interaction with a semi-major axis greater than the semi-major axis of exact corotation, 
whereas body 4 begins with a semi-major axis less than that of exact corotation.
Since body 3 is losing angular momentum and energy, its semi-major axis decreases, and it moves towards exact corotation. 
Since body 4 is gaining angular momentum and energy, its semi-major axis increases, so it also moves towards exact corotation. 

As we can see in these examples, then, it is the combination of the epicyclic phase of the encounters and their locations in phase space that determines
which direction the bodies move relative to exact corotation.  

In all of these encounters, we can see that energy is transferred from one body to the other, but total energy is conserved. 
Thus we can see that there is a difference between this energy and the energy maxima usually defined for corotation resonances
\citep{1986AJ.....92..490G, 1991Icar...89..197S, 1995netr.conf..703P, 2002Natur.417...45N}. 
This is because the classical picture of corotation resonances involves averaging over many terms in the potential, 
while these encounters occur over a short timescale where those terms in the potential cannot be ignored.
This means that close encounters between bodies within a corotation resonance will disperse particles 
in phase space in a manner that is largely independent of the corotation sites.

This basic finding has important implications for the stability of ring arcs, 
particularly dense arcs like those found in Neptune's rings, 
where inter-particle collisions should be common, and inelastic interactions such as accretion can occur. 
On the one hand, such interactions could be more apt to disperse material out of the stable corotation sites. 
On the other hand, dissipative collisions might not necessarily require material to move away from the exact corotation sites. 
Detailed numerical simulations of collisional ring arcs will likely be needed to properly investigate these issues.

\acknowledgements
We are grateful to many individuals for useful discussions, especially R. Chancia, J. Ahlers, and D. Hull-Nye.
We thank NASA for the support through the Cassini Data Analysis and Participating Scientist Program grant NNX15AQ67G.
We also thank an anonymous reviewer for helpful comments. 

\bibliography{RingArcPaperv13} 
\bibliographystyle{aasjournal}

\end{document}